\documentclass[12pt]{article}
\usepackage{graphicx}
\DeclareGraphicsExtensions{.pdf}
\usepackage{float} 
\usepackage{amsmath}
\usepackage{slashed}
\usepackage{amsfonts}   
\usepackage{amssymb}

\begin{document}
\date{}
\title{{\bf{\Large Decoding the Spin-Matrix limit of strings on $AdS_5 \times S^5$}}}
\author{
 {\bf {\normalsize Dibakar Roychowdhury}$
$\thanks{E-mail:  dibakarphys@gmail.com, dibakar.roychowdhury@ph.iitr.ac.in}}\\
 {\normalsize  Department of Physics, Indian Institute of Technology Roorkee,}\\
  {\normalsize Roorkee 247667, Uttarakhand, India}
\\[0.3cm]
}

\maketitle
\begin{abstract}
In this Letter, we explore nonrelativistic string solutions in various subsectors of the $ SU(1,2|3) $ SMT strings that correspond to different spin groups and satisfy the respective BPS bounds. In particular, we carry out an explicit analysis on rotating string solutions in the light of recently proposed SMT limits. We explore newly constructed SMT limits of type IIB (super) strings on $ AdS_5 \times S^5 $ and estimate the corresponding leading order stringy corrections near the respective BPS bounds.
\end{abstract}
\section{Overview and Motivation}
The conjectured duality between various near BPS corner(s) of $\mathcal{N}=4$ SYM \cite{Harmark:2006di}-\cite{Harmark:2008gm} and the stringy dynamics over non-Lorentzian manifolds has drawn lots of attention in the recent years. Major breakthrough will come once we understand these nonrelativistic string states in terms of the dual quantum mechanical degrees of freedom known as the Spin-Matrix theory (SMT) \cite{Harmark:2014mpa}-\cite{Baiguera:2020mgk}. The first step is therefore to strengthen our current understanding in the strong coupling regime of the correspondence and in particular to find an interpretation for the propagating (nonrelativistic) string states (on $ AdS_5 \times S^5 $) in terms of dual SMT degrees of freedom in the near BPS limit(s) of $\mathcal{N}=4$ SYM on $ R \times S^3 $.

Till date there have been two parallel formulations of nonrelativistic sigma models \cite{Gomis:2000bd}-\cite{Gomis:2005pg} over non-Lorentzian geometries. The corresponding target space manifolds have been categorised as string Newton-Cartan (SNC) \cite{Bergshoeff:2018yvt}-\cite{Bergshoeff:2019pij} and torsional Newton-Cartan (TNC) geometries \cite{Harmark:2017rpg}-\cite{Roychowdhury:2020kma}. 

TNC geometries are of particular interest, as the large $c$ limit of TNC string sigma models is conjectured to be dual to SMT limits of $\mathcal{N}=4$ SYM on $R \times S^3$ \cite{Harmark:2017rpg}-\cite{Harmark:2018cdl}. The corresponding target space over which these nonrelativistic strings propagate are categorised as $U(1)$ Galilean geometries. To understand the conjectured duality, it is therefore crucial to unveil the nonrelativistic string spectra over $U(1)$ Galilean manifolds those correspond to different near BPS corners of $\mathcal{N}=4$ SYM on $R \times S^3$.  

The purpose of this paper is to carry out calculations in the light of recently proposed SMT limits \cite{Harmark:2020vll} and perform an extensive analysis on semiclassical string solutions corresponding to different subsectors of the full $ PSU(1,2|3) $ group. We start by considering $ SU(2|3) $ SMT limit of strings on $ AdS_5\times S^5 $ and consider various special limits within it. 

The first non trivial example we focus on is the $ SU(2)\subset SU(2|3)  $  subsector where we consider two different string embeddings. In the first example, the string soliton is considered to be stretched along the polar coordinate ($ \theta $) of $ S^2 $ while simultaneously wrapping its isometry direction ($ \varphi $). Given a particular value of the winding mode namely, $ N=2 $ we obtain a nice dispersion relation that can be identified with magnon excitations along the $ SU(2) $ spin chain in the SMT limit. 

In the second example, we consider a configuration in which the string is stretched along $ \theta $ only while its end points are rotating along $ \varphi $. The resulting dispersion relation is obtained in terms of special elliptic functions. It would be really nice to see these results from the perspective of a dual spin chain model.

Like in the $ SU(2) $ case, we consider similar string embeddings for the full $ SU(2|3) $ SMT limit. The spectrum that we obtain essentially confirms similar qualitative features those have already been observed for the $ SU(2) $ example. 

The next example that we consider is that of $ SU(1,1) $ SMT limit in which we turn on one spin along $ S^3 \subset AdS_5 $ and one R-charge along $ S^5 $. This is the limit in which SMT strings describe the strong coupling dynamics near the BPS bound $ Q=S_1+J_1 $ in $ \mathcal{N}=4 $ SYM. Our analysis reveals that for frequencies $ \nu \geq \frac{N^2}{2} $ the dispersion relation is of the form, $ H-Q \sim J \sqrt{\nu} $.

We extend our analysis further for the $ SU(1,1|2) $ as well as $ SU(1,2|3) $ SMT limits of strings on $ AdS_5 \times S^5 $. These are the near BPS corners of $ \mathcal{N}=4 $ SYM with charges $ Q=S_1+J_1 +J_2 $ and $ Q=S_1 + S_2 +J_1 $ respectively. In order to probe the corresponding strong coupling SMT dynamics, we allow the string soliton to wrap as well as spin along the $ w $ circle extended both in the $ AdS_5 $ as well as $ S^5 $ factors of the geometry \cite{Harmark:2020vll}. This sources the combination of charges $ Q_1=-i\partial_{w}=S_1 +J_1 $ in the BPS limit of the full superconformal algebra. 

In the $ SU(1,1|2) $ SMT example, the remaining R-charge ($ J_2 $) comes from the rotation along the isometry of $ S^2\subset S^5 $. On the other hand, in case of $ SU(1,2|2) $ SMT limit, the remaining charge ($ S_2 $) arises due to rotation along $ S^3 \subset AdS_5 $.

We conclude our analysis with a particular focus towards the near BPS corner $Q=S_1+S_2+J_1+J_2+J_3$ of $\mathcal{N}=4$ SYM which corresponds to the full $PSU(1,2|3)$ SMT limit of strings on $AdS_5 \times S^5$. This is the limit in which we turn on spins along all the isometries of the corresponding $U(1)$ Galilean target space. The rotation along $w$ circle gives rise to the charge combination $Q_1=-i\partial_{w}=S_1 + J_1$ as usual. The rest of the charges ($S_2$, $J_2$ and $J_3$) are associated with the rotation of the string along the remaining three isometries of the target spacetime \cite{Harmark:2020vll}.

We draw our conclusion in Section 3 along with some future remarks.
\section{Rotating string solutions}
\subsection{$ SU(2|3) $ SMT strings}
We now move on towards discussing various semi-classical string solutions in the light of recently proposed SMT limits \cite{Harmark:2020vll}. We start by discussing the $ SU(2|3) $ sector of $ \mathcal{N}=4 $ and zoom into its near BPS limit. The dual (flat world-sheet) gauge fixed sigma model Lagrangian may be schematically expressed as,
\begin{eqnarray}
\mathcal{L}_P = -(\sin^{2}\xi -\frac{1}{2})\dot{\psi}-\frac{1}{2}\sin^{2}\xi \cos\theta \dot{\varphi}+\frac{\xi'^2}{2}+\frac{1}{8}\sin^2\xi \theta'^2\nonumber\\
+\frac{1}{8}\sin^2\xi (\sin^2\theta + \cos^2\theta \cos^2\xi)\varphi'^2 +\frac{1}{2}\sin^2\xi \cos^2\xi \psi'^2 +\frac{1}{2}\cos\theta \sin^2\xi \cos^2\xi \psi' \varphi'.
\end{eqnarray}

The corresponding charges associated with the sigma model are given by, $ Q_1=-i\partial_{\psi}\sim \frac{1}{2}(J_1+J_2 -J_3) $ and $ Q_2 =-i\partial_{\varphi}\sim \frac{1}{2}(J_2-J_1) $ where $ J_i (i=1,2,3) $ are the original R-charges associated to $ S^5 $ \cite{Harmark:2020vll}. 

To proceed further, we choose to work with the string embedding of the following form in which the string wraps as well as rotates along the isometry directions ($ \psi $ and $ \varphi $) of the target space,
\begin{eqnarray}
\xi = \xi (\sigma^1)~;~\theta = \theta (\sigma^1)~;~\psi = \omega_1 \sigma^0 + M \sigma^1 ~;~\varphi = \omega_2 \sigma^0 + N \sigma^1
\end{eqnarray}

This yields the equations of motion of the following from
\begin{eqnarray}
\xi'' +\sin\xi \cos\xi (2\omega_1 + \omega_2 \cos\theta)-\frac{1}{4}\sin\xi \cos\xi \theta'^2 - \frac{N^2}{4}\sin\xi \cos\xi (\sin^2\theta + \cos^2\theta \cos^2\xi)\nonumber\\
-\frac{N^2}{4}\sin^3\xi \cos\xi \cos^2\theta -M \sin\xi \cos\xi (1-2 \sin^2\xi)(M + N \cos\theta)=0,
\label{e74}
\end{eqnarray}
and
\begin{eqnarray}
\theta'' +2 \cot\xi \xi'\theta' -2\omega_2 \sin\theta -N^2 \sin^2\xi \sin\theta \cos\theta +2MN \sin\theta \cos^2\xi =0.
\label{e75}
\end{eqnarray}

To solve (\ref{e74})-(\ref{e75}) and in order to obtain the corresponding semiclassical string spectra below we consider the following limiting cases.\\\\
$ \bullet $ \textbf{$  SU(2)$ SMT limit:} (i) The $ SU(2) $ SMT limit is achieved by setting $ \xi = \frac{\pi}{2} $ and $ \psi =  $ fixed  \cite{Harmark:2020vll} which for the present case yield the following equation,
\begin{eqnarray}
\theta'' -2\omega_2 \sin\theta -N^2  \sin\theta \cos\theta  =0.
\label{e76}
\end{eqnarray}

Integrating (\ref{e76}) once we find,
\begin{eqnarray}
\theta' (\sigma^1)=2N\sqrt{\left( \sin^2\alpha -\sin^2\frac{\theta}{2}\right) \left( \sin^2\frac{\theta}{2}-\sin^2\beta \right) }
\label{e77}
\end{eqnarray}
where,
\begin{eqnarray}
 \sin^2\alpha + \sin^2\beta &=&1+\frac{2\omega_2}{N^2}\\
 \sin^2\alpha \sin^2\beta &=&\frac{1}{4}+\frac{\omega_2}{N^2}-\frac{C}{2N^2}.
\end{eqnarray}

This leads to the following dispersion relation associated with the $  SU(2)$ SMT strings,
\begin{eqnarray}
(H-Q_{2})_{reg}=\frac{J}{16\pi}\Delta\varphi_m +\frac{J}{8\pi}(2-\log 2)+\delta_{ct}
\label{e80}
\end{eqnarray}
where $Q_{2} \sim J_{\varphi}\sim -i\partial_{\varphi}$ is the R-charge corresponding to the rotation along $ \varphi $ and $ J $ is the winding parameter that corresponds to the length of the spin chain in the planar limit of SMT \cite{Harmark:2020vll} and $ \delta_{ct}=-\frac{J}{4\pi}\log\epsilon $ is the counter term. Furthermore, we set the winding number $ N=2 $ together with the fact $\alpha=\frac{\pi}{2} $ and $\beta \sim \epsilon \sim 0 $. 

The angular difference between the end points of the soliton,
\begin{eqnarray}
\Delta\varphi_m =4 \int_{\epsilon}^{\pi/2}\frac{d\theta}{\theta'}
\end{eqnarray}
should be identified with the magnon excitation propagating with some momentum $ p_m $ along the $ SU(2) $ spin chain. 

To see this explicitly, we consider the magnon dispersion relation of the form \cite{Harmark:2018cdl}
\begin{eqnarray}
H-Q = \sqrt{1+\frac{\lambda_{SYM}}{\pi^2}\sin^2\frac{p_m}{2}}-1
\end{eqnarray}
and replace the SYM coupling with SMT coupling as $ \lambda_{SYM} = \frac{\mathfrak{g}_{SMT}}{c^2} $.  Considering the fact, $ p_m \ll 1 $ we obtain the dispersion relation in the form,
\begin{eqnarray}
H-J \sim \frac{\mathfrak{g}_{SMT}}{8\pi^2 c^2}p^2_m +\cdots
\end{eqnarray}
which reveals the fact that,
\begin{eqnarray}
\Delta\varphi_m \sim \frac{\mathfrak{g}_{SMT}}{J}p^2_m.
\label{e84}
\end{eqnarray}

(ii) The second example that we consider is that of a rotating string that is stretched along $ \theta $ only with its end points rotating along $ \varphi $. This corresponds to setting $ N=0 $ in (\ref{e76}) which thereby yields,
\begin{eqnarray}
\theta'(\sigma^1)=\sqrt{2C - 4 \omega_2 \cos\theta}
\label{e85}
\end{eqnarray}
where $ C $ is the constant of integration.

Notice that, here $ \theta \in S^2 $ is the polar angle along which the string is stretched. Imagine that the string is extended through the north pole on both sides such that its end points lie on the equator namely, $ -\frac{\pi}{2}\leq \theta \leq \frac{\pi}{2} $. Clearly, in this picture, $ C=2 \omega_2 $ corresponds to the turning point $ \theta_0 =0 $. In other words, the constant $ C $ can be fixed in terms of the geometrical angle $ \theta_0 = \arccos(\frac{C}{2\omega_2}) $ in the bulk.

Using (\ref{e85}) and setting $ C=2\omega_2 $, the spectrum turns out to be
\begin{eqnarray}
(H-Q_2)_{reg} &=& \frac{J}{4\pi}\int_{0}^{\pi/2}\frac{ \omega_2 - (\omega_2 +1) \cos \theta}{\sqrt{\omega_{2} -\omega_2  \cos\theta}}+\delta_{ct}\nonumber\\
&=&-\frac{J}{4\pi}\left(\frac{4 \omega_2 +4+\sqrt{2} \log \left(3-2 \sqrt{2}\right)}{\sqrt{\omega_2 }}+\frac{4\sqrt{2} \left( \omega_2 +1 - \log2 \right)}{\sqrt{\omega_2 }} \right) +\delta_{ct}
\end{eqnarray}
where the counter term is given by, $ \delta_{ct} = \frac{J}{\sqrt{2 \omega_2}\pi}\log\epsilon$.\\

$ \bullet $ \textbf{$  SU(2|3)$ SMT limit of rotating strings:} (i) We now return to the original set of equations (\ref{e74}) and (\ref{e75}) and set $ M=N=0 $. This simplifies the dynamics as,
\begin{eqnarray}
\label{e87}
\xi'' +\sin\xi \cos\xi (2\omega_1 + \omega_2 \cos\theta)-\frac{1}{4}\sin\xi \cos\xi \theta'^2 &=&0\\
\theta'' +2 \cot\xi \xi'\theta' -2\omega_2 \sin\theta & =&0.
\label{e88}
\end{eqnarray}

In order to simplify the dynamics further, we place the string soliton at the north pole ($ \theta=0 $) of $ S^2 \subset S^5 $ and consider to be stretched along $ \xi $ only. This is a consistent choice to start with as this clearly solves (\ref{e88}) and resides within the $  SU(2|3)$ SMT limit.

Substituting this ansatz into (\ref{e87}) we find,
\begin{eqnarray}
\xi' (\sigma^1)=\sqrt{C-\Omega \sin^2\xi}=\sqrt{\Omega}\sqrt{\sin^2\xi_{max}-\sin^2\xi}
\end{eqnarray}
where $ \Omega = \omega_2 +2 \omega_1 $ denotes the sum of angular frequencies.

Notice that, here $ \xi \in S^3 $ is the polar angle along which the string is extended. Clearly, for the given string embedding, $ 0 \leq \xi \leq \xi_{max} $. In other words, the string can be stretched upto the maximal angle, $ \xi_{max}=\arcsin\sqrt{\frac{C}{\Omega}} $.  Using this language, the integration constant $ C $ can be related to the total angular frequency of the system which in turn is completely determined in terms of the maximal (geometrical) angle $ \xi_{max} $ in the bulk.

The dispersion relation associated with $  SU(2|3)$ SMT sigma model turns out to be,
\begin{eqnarray}
\label{eqn19}
(H -Q)_{reg}&=&\frac{J}{4\pi}\int_{0}^{\pi/2}d\xi \frac{\Omega +1-(\Omega +3)\sin^2\xi}{\sqrt{\Omega -\Omega \sin^2\xi}}+\delta_{ct}\nonumber\\
&=&-\frac{J}{4\pi}\frac{\Omega +3-2\log 2}{\sqrt{\Omega}}+\delta_{ct}
\end{eqnarray}
where we set, $ C=\Omega $ together with, $ Q \sim Q_1+Q_2$ which denotes the total R-charge in transformed coordinates \cite{Harmark:2020vll}. Moreover, here $ \delta_{ct}=\frac{J}{2\pi\sqrt{\Omega}}\log\epsilon $ is the counter term that is needed in order to tame the divergence near $ \xi \sim \xi_{max}\sim \frac{\pi}{2} $.

(ii) In the second example, we generalise our results by turning on both the winding numbers namely we consider $ N\neq 0 $ and $ M \neq 0 $ and set $ \theta =0 $ as before. Like before, (\ref{e75}) is automatically satisfied whereas on the other hand, from (\ref{e74}) we find,
\begin{eqnarray}
\xi'(\sigma^1)= \sqrt{M(M+N)}\sqrt{(\sin^2\xi_{max}-\sin^2\xi)(\sin^2\xi -\sin^2\xi_{min})}
\end{eqnarray}
where we identify,
\begin{eqnarray}
\sin^2\xi_{max} +\sin^2\xi_{min}&=&\frac{\frac{N^2}{4}-2\omega_1 - \omega_2 +M(M+N)}{M(M+N)}\\
\sin^2\xi_{max} \sin^2\xi_{min}&=&\frac{\frac{N^2}{8}-C}{M(M+N)}.
\end{eqnarray}

This finally leads to the conserved charges associated with the string motion as,
\begin{eqnarray}
H&=&J\frac{8 M (M+N)+N^2}{32 \pi \sqrt{M (M+N)}}\\
J_{\psi}&=&\frac{J}{\pi}(\frac{\Delta \phi_m}{4}+\frac{\log \left(\frac{\epsilon}{2}\right)}{ \sqrt{M (M+N)}})\\
J_{\varphi}&=&\frac{J}{\pi}(\frac{\Delta \phi_m}{4}+\frac{\log \left(\frac{\epsilon}{2}\right)}{2  \sqrt{M (M+N)}})
\end{eqnarray}
where we evaluate the integrals by setting $ \xi_{max}=\frac{\pi}{2} $ and $ \xi_{min}\sim \epsilon \sim 0 $. As before, $ \Delta \phi_m \sim p^2_m $ should be identified with the (square of the) magnon momentum (\ref{e84}) along the $ SU(2|3) $ spin chain leading to the dispersion relation of the following form,
\begin{eqnarray}
(H-|Q|)_{reg}\sim \frac{ J}{2\pi} \Delta\phi_m +\cdots
\end{eqnarray}
\subsection{$ SU(1,1) $ SMT strings}
We now turn our attention towards the near BPS corner $ Q=-i\partial_{w}=S_1+J_1 $ of $ \mathcal{N}=4 $ SYM and compute the corresponding spectrum at strong coupling. The states in the dual SMT have an interpretation in terms of nonrelativistic strings described by \cite{Harmark:2020vll},
\begin{eqnarray}
\mathcal{L}_P=\sinh^2\rho \dot{w}-\frac{1}{2}(\rho'^2 + \sinh^2\rho \cosh^2 \rho w'^2).
\end{eqnarray}

In order to evaluate the stringy dynamics near this BPS bound we choose a \emph{folded} spinning string embedding of the following form,
\begin{eqnarray}
\rho = \rho (\sigma^1)~;~w= \nu \sigma^0 + N \sigma^1
\end{eqnarray}
which yields an equation of motion of the following form,
\begin{eqnarray}
\rho'' +2 \nu \sinh\rho \cosh\rho - N^2 \sinh\rho \cosh\rho (1+ 2 \sinh^2\rho)=0.
\label{e100}
\end{eqnarray}

(i) We first consider the case in which $ \nu> \frac{N^2}{2} $. In this parameter range the above equation (\ref{e100}) can be integrated once to obtain,
\begin{eqnarray}
\rho'(\sigma^1)=N\sqrt{(\sinh^2\rho - \sinh^2\rho_{+})(\sinh^2\rho - \sinh^2\rho_{-})}
\label{e101}
\end{eqnarray}
where we identify,
\begin{eqnarray}
\sinh^2\rho_{\pm}=\frac{1}{2}\left( \frac{2\nu}{N^2}-1\right) \pm \frac{1}{2}\sqrt{\left( \frac{2 \nu}{N^2}-1\right) ^2 -\frac{4C^2}{N^2}}
\end{eqnarray}
while $ C $ being the constant of integration. To proceed further, without any loss of generality we set, $ C=0 $ which thereby sets the lower limit as $ \rho_{-}=0 $.

The dispersion relation associated with $ SU(1,1) $ SMT strings turns out to be,
\begin{eqnarray}
H-Q&=&\frac{2 N J}{\pi}\int_0^{\rho_+}d\rho \sinh\rho \sqrt{\cosh^2\rho -\frac{2\nu}{N^2}}\nonumber\\
&=&-\frac{ J}{\pi N}\left( N^2 \sqrt{\frac{2 \nu }{N^2}-1}+\nu  \log \left(\frac{2 \nu }{N^2}\right)-2 \nu  \log \left(\sqrt{\frac{2 \nu }{N^2}-1}+1\right)\right).
\label{e103}
\end{eqnarray}

Let us consider the \emph{long} string limit in which $ \nu \gg \frac{N^2}{2} $ which simplifies (\ref{e103}) as,
\begin{eqnarray}
\label{eqn33}
H-Q \sim \frac{\sqrt{2}J}{\pi }\sqrt{\nu}.
\end{eqnarray}

(ii) In the second example, we consider $ \nu = \frac{N^2}{2} $ which yields
\begin{eqnarray}
\label{eqn34}
\rho'(\sigma^1)=N\sqrt{\Big |\sinh^4\rho -\frac{C^2}{N^2}\Big |}=N\sqrt{\sinh^4\rho_m -\sinh^4\rho},
\end{eqnarray} 
where, $ \rho_m = \sinh^{-1}\sqrt{\frac{C}{N}} $ is the maximum radial distance to which the string is extended. In other words, the constant of integration ($ C $) is completely determined in terms of the maximal radial distance ($ \rho_m $) that the string can be stretched inside the bulk. Clearly, in the above picture one should identify $ \rho =\rho_m $ as the turning point.

Using (\ref{eqn34}), the spectrum may be obtained as
\begin{eqnarray}
H-Q&\sim &\frac{JN}{\pi}\int_{0}^{\rho_m}\frac{\sinh^4\rho_m}{\sqrt{\sinh^4\rho_m -\sinh^4\rho}}\nonumber\\
& \sim &- \frac{JN}{\pi} \tanh\rho_m \sinh^2\rho_m F\left(-\frac{\pi}{2} |1-2 \text{sech}^2\rho_m\right),
\label{e106}
\end{eqnarray}
where we apply L Hospital's rule in order to estimate the integral (\ref{e106}) near the upper bound, $ \rho \sim \rho_m $.

(iii) In the third example, we consider the parameter range $ \nu <\frac{N^2}{2} $. The corresponding solution to (\ref{e100}) turns out to be,
\begin{eqnarray}
\rho'(\sigma^1)=N\sqrt{\sinh^4\rho + \beta_c \sinh^2\rho - \frac{C^2}{N^2}}
\end{eqnarray}
where $ \beta_c = 1-\frac{2\nu}{N^2} $.

Corresponding to turning point(s) one must set, $ \rho'=0 $ which thereby yields one real positive root ($ \rho_m $) satisfying,
\begin{eqnarray}
\sinh^2\rho_m = \frac{1}{2}\left(\sqrt{\beta^2_c + \frac{4C^2}{N^2} }-\beta_c \right). 
\end{eqnarray}

This leads to the following dispersion relation in the near BPS limit of $ \mathcal{N}=4 $ SYM,
\begin{eqnarray}
H -Q = \frac{2JN}{\pi}\int_{0}^{\rho_m} \frac{\beta_{c}  \sinh ^2\rho -\frac{C^2}{2N^2}+\sinh ^4\rho }{\sqrt{\beta_{c}  \sinh ^2\rho -\frac{C^2}{N^2}+\sinh ^4\rho }} \, d\rho.
\label{109}
\end{eqnarray}

Considering the \emph{short} string limit $ \frac{C}{N}\ll 1 $, the above integral (\ref{109}) reduces to
\begin{eqnarray}
\label{eqn39}
H -Q \sim \frac{JN}{\pi} \left(\beta_{c} ^{3/2}-(\beta_{c} +1) \beta_{c}  \log \left(\sqrt{2} \left(\sqrt{\beta_{c} }+1\right)\right)+2 \log \left(\sqrt{\beta_{c} }+1\right)+\log 2\right).
\end{eqnarray}
\subsection{$ SU(1,1|2) $ SMT strings}
The $ SU(1,1|2) $ SMT strings are described by the sigma model of the form \cite{Harmark:2020vll},
\begin{eqnarray}
\mathcal{L}_P = 2\sinh^2\rho \dot{w}+\cos\theta \dot{\varphi}-\rho'^2 -\sinh^2\rho \cosh^2\rho \omega'^2 - \frac{1}{4}(\theta'^2 + \sin^2\theta \varphi'^2)
\end{eqnarray}
which essentially represents states near the BPS bound $ Q=S_1+J_1+J_2 \equiv Q_1 + Q_2$ where $ Q_1= -i\partial_{w}=S_1+J_1  $ in the Fubini-Study coordinates \cite{Harmark:2020vll}. On the other hand, the remaining R-charge $Q_2= -i\partial_{\varphi}=J_2 $ comes from the two sphere  $ S^2 \subset S^5 $.

To proceed further, we propose a string embedding of the following form
\begin{eqnarray}
w = \nu_1 \sigma^0 + N \sigma^1 ~;~\varphi = \nu_2 \sigma^0 + M \sigma^1 ~;~\theta = \theta (\sigma^1)~;~\rho = \rho (\sigma^1).
\end{eqnarray}

The resulting equations of motion turn out to be,
\begin{eqnarray}
\label{e113}
\rho'' +(2\nu_1 - N^2)\sinh\rho \cosh\rho -2N^2 \sinh^3\rho \cosh\rho &=& 0\\
\theta'' -2 \nu_2 \sin\theta - M^2 \sin\theta \cos\theta &=&0.
\label{e114}
\end{eqnarray}

Integrating the above equations once we find,
\begin{eqnarray}
\rho'(\sigma^1)&=&N\sqrt{(\sinh^2\rho - \sinh^2\rho_{+})(\sinh^2\rho - \sinh^2\rho_{-})}\\
\theta' (\sigma^1)&=&2M\sqrt{\left( \sin^2\alpha -\sin^2\frac{\theta}{2}\right) \left( \sin^2\frac{\theta}{2}-\sin^2\beta \right) }
\end{eqnarray}
which have already been found previously in (\ref{e77}) and (\ref{e101}).

Setting $ \alpha = \frac{\pi}{2} $ and $ \rho_- = 0 $ this further yields,
\begin{eqnarray}
\label{eqn46}
(H - Q)_{reg} \sim \frac{J}{\pi }\left( \sqrt{2\nu_1}+\frac{\nu_2}{2M}\log2 +\frac{M}{4}\right) +\delta_{ct}
\end{eqnarray}
where we add a counter term $\delta_{ct} \sim -\frac{\nu_2}{2M}\log \epsilon$ in addition to the fact that $ \nu_1 \gg \frac{N^2}{2} $.
\subsection{$ SU(1,2|2) $ SMT strings}
The next example we consider is that of $ SU(1,2|2) $ SMT strings which basically describe dynamics in the near BPS limit $ Q=Q_1+Q_2=S_1 +S_2 +J_1 $ of $ \mathcal{N}=4 $ SYM. Here $ S_1 $ and $ S_2 $ are the spins along $ S^3 \subset AdS_5 $ and $ J_1 $ is the R-charge associated to rotation in $ S^5 $. 

The corresponding dual sigma model Lagrangian turns out to be  \cite{Harmark:2020vll},
\begin{eqnarray}
\mathcal{L}_P=2\sinh^2 \rho \dot{w}+\sinh^2\rho \cos\bar{\theta}\dot{\bar{\varphi}}-\rho'^2 - \frac{1}{4}\sinh^2\rho (\bar{\theta}'^2 + \sin^2\bar{\theta}\bar{\varphi}'^2)\nonumber\\
-\sinh^2\rho \cosh^2\rho (w'+\frac{1}{2}\cos\bar{\theta}\bar{\varphi}')^2.
\end{eqnarray}  

Like in the previous example, the $ w $ circle belongs to both $ AdS_5 $ and $ S^5 $ which thereby yields the fact that, $ Q_1 = -i\partial_{w}=S_1+J_1 $. On the other hand, $ \bar{\theta} $ and $ \bar{\varphi} $ belongs to the three sphere $ S^3\subset AdS_5 $. In particular, $ \bar{\varphi} $ is an isometry which produces the second conserved charge $ Q_2 = -i\partial_{\bar{\varphi}}=S_2 $. 

To proceed further, we choose to work with the string embedding of the following form,
\begin{eqnarray}
\rho = \rho (\sigma^1)~;~\bar{\theta}=\bar{\theta}(\sigma^1)~;~w = \nu_1 \sigma^0 + N \sigma^1 ~;~\bar{\varphi}=\nu_2 \sigma^0 + M \sigma^1.
\end{eqnarray}

The resulting equations of motion turn out to be
\begin{eqnarray}
\label{e120}
\rho'' +\sinh\rho \cosh\rho (2\nu_1 +\nu_2 \cos\bar{\theta})-\frac{1}{4}\sinh\rho \cosh\rho (\bar{\theta}'^2 +M^2  \sin^2\bar{\theta})\nonumber\\
-\sinh\rho \cosh\rho (1+2\sinh^2\rho)(N+\frac{M}{2}\cos\bar{\theta})^2=0,\\
 \bar{\theta}''-2\nu_2  \sin\bar{\theta}+2\coth\rho \rho' \bar{\theta}' -\frac{M^2}{2} \sin2\bar{\theta}+2M\cosh^2\rho (N+\frac{M}{2}\cos\bar{\theta})\sin\bar{\theta}=0.
 \label{e121}
\end{eqnarray}

Clearly, the choice $ \bar{\theta}=0 $ solves the corresponding equation of motion (\ref{e121}). Following this choice, we essentially restrict ourselves to some particular submanifold in Fubini-Study coordinates \cite{Harmark:2020vll}. Substituting this into (\ref{e120}) we find,
\begin{eqnarray}
\rho'(\sigma^1)=\gamma \sqrt{(\sinh^2\rho - \sinh^2\rho_{+})(\sinh^2\rho - \sinh^2\rho_{-})}
\end{eqnarray}
where we find the rots to be,
\begin{eqnarray}
\sinh^2\rho_{\pm}=\frac{1}{2}\left( \frac{\kappa^2}{\gamma^2}\pm \sqrt{\frac{\kappa^4}{\gamma^4}-\frac{4C^2}{\gamma^2}}\right) 
\end{eqnarray}
where we define, $ \gamma^2 = (N +\frac{M}{2})^2 $ and $ \kappa=\sqrt{2\nu_1 + \nu_2 -\gamma^2} $.

In the process of obtaining the corresponding spectrum we set, $ C=0 $ which therefore fixes one of the roots $ \rho_-=0 $. The other end where the string folds corresponds to $ \sinh^2\rho_+ = \frac{\kappa^2}{\gamma^2} $. Considering these facts, the associated dispersion relation turns out to be,
\begin{eqnarray}
\label{eqn53}
H-Q &=&\gamma (1+\gamma^2)\int_0^{\rho_+}d\rho \sinh\rho \sqrt{\sinh^2\rho - \frac{\kappa^2}{\gamma^2}}\nonumber\\
&\approx &\frac{J}{2\pi}\left( \frac{\kappa}{\gamma}\right)^6 \left( \gamma - \left( \frac{\kappa}{\gamma}\right)^{-4}\log \left( \frac{2\kappa}{\gamma}\right)+\cdots \right)  
\end{eqnarray}
where we set the long string limit namely, $ |\frac{\kappa}{\gamma}|\gg 1 $.
\subsection{$ PSU(1,2|3) $ SMT strings}
Finally, we extend our analysis for the full $ PSU(1,2|3) $ SMT limit on $ AdS_5 \times S^5 $ which corresponds to the near BPS corner $ Q=Q_1+Q_2 +Q_3+Q_4 =S_1+S_2+J_1+J_2+J_3$ of $ \mathcal{N}=4 $ SYM. The dual non relativistic sigma model Lagrangian is given by \cite{Harmark:2020vll},
\begin{eqnarray}
\label{eqn54}
\mathcal{L}_P = 2\sinh^2\rho \dot{w}+\sinh^2\rho \cos\bar{\theta}\dot{\bar{\varphi}}-\cos(2\xi) \dot{\psi}+\sin^2\xi \cos\theta \dot{\varphi}-\rho'^2 -\xi'^2\nonumber\\
-\frac{1}{4}\sinh^2\rho \bar{\theta}'^2 - \frac{1}{4}\sin^2\xi \theta'^2
\end{eqnarray}
where we consider that the end points of the string soliton to be rotating along the isometires of the spacetime rather than wrapping around it.

Notice that, the above Lagrangian (\ref{eqn54}) is a special case of \cite{Harmark:2020vll} where we set, $ w'=\psi'=\varphi' =\bar{\varphi}'=0 $. In other words, the string is considered to have zero winding modes along these isometry directions. Given this set up, one can in fact reproduce various SMT string subsectors (those obtained previously) as a special case of $ PSU(1,2|3) $ SMT strings. We discuss these limiting cases in detail below (\ref{e133}).

Here $ \psi $ and $ \varphi $ are the coordinates associated to $ S^5 $. On the other hand, $ \bar{\varphi} $ belongs to $ S^3 \subset AdS_5 $. The charges associated to different isometries are enumerated below,
\begin{eqnarray}
Q_1 &=& -i\partial_{w}=S_1 + J_1~;~Q_2 = -i\partial_{\bar{\varphi}}=S_2 \\
Q_3 &=&-i\partial_{\psi}=J_2~;~Q_4=-i\partial_{\varphi}=J_3.
\end{eqnarray}

To proceed further, we choose to work with a stringy ansatz of the following form,
\begin{eqnarray}
\rho = \rho (\sigma^1)~;~\xi = \xi (\sigma^1),~;~w = \nu_1 \sigma^0 ~;~ \bar{\varphi}= \nu_2 \sigma^0 \\
\bar{\theta}= \bar{\theta}(\sigma^1)~;~\theta = \theta (\sigma^1)~;~\psi = \nu_3 \sigma^0 ~;~ \varphi =\nu_4 \sigma^0.
\end{eqnarray}

The resulting equations of motion are given by
\begin{eqnarray}
\label{e130}
\rho'' +\sinh\rho \cosh\rho (2 \nu_1 + \nu_2 \cos\bar{\theta})-\frac{1}{4}\sinh\rho \cosh\rho \bar{\theta}'^2 &=&0,\\
\label{e131}
\xi'' + (2\nu_3 +\nu_4 \cos\theta )\sin\xi \cos\xi - \frac{\theta'^2}{4}\sin\xi \cos\xi &=&0,\\
\label{e132}
\bar{\theta}'' -2\nu_2 \sin\bar{\theta}+ 2\coth\rho \rho' \bar{\theta}' &=&0,\\
\theta'' -2\nu_4  \sin\theta + 2\cot\xi \xi' \theta' &=&0.
\label{e133}
\end{eqnarray}

As mentioned above, the above set of equations (\ref{e130})-(\ref{e133}) captures the SMT string dynamics in various SMT sub-sectors those discussed previously. For example, the $ SU(2) $ subsector (\ref{e76}) can be recovered (as a special limiting case) by setting $ \rho =0=\bar{\theta} $ , $ \xi = \frac{\pi}{2} $ together with $ N=\varphi' =0 $. Here, one has to identify $ \omega_2 = \nu_4 $. Following similar steps, one can in fact identify the bigger $ SU(2|3) $ subgroup (\ref{e87})-(\ref{e88}) by setting $ \rho =0=\bar{\theta} $. 

On the other hand, the $ SU(1,1) $ SMT strings (\ref{e100}) are recovered in the limit in which one switches off all the coordinates together with, $ \nu_2 =0 $ as well as $ N=w'=0 $. In a similar spirit, $ SU(1,1|2) $ SMT strings (\ref{e113})-(\ref{e114}) are recovered by setting, $ M=\varphi' =0 $ together with $ \nu_2=0 $ , $ \xi =\frac{\pi}{2} $ , $ \bar{\theta}=0 $ and $ N=w'=0 $. On a similar note, one can easily identify the $ SU(1,2|2) $ SMT strings (\ref{e120})-(\ref{e121}) by switching off $ \theta $ and turning on $ \bar{\theta} $.

Like in the previous example, one can choose $ \bar{\theta}=\theta =0 $ as a solution as these solve (\ref{e132}) and (\ref{e133}) and preserve the charges as well. However, with these choices one essentially restricts to the sub-manifold parametrized by the Fubini-Study coordinates $\lbrace z_0, z_2, w_2, w_3\rbrace  $ \cite{Harmark:2020vll}. Substituting these solutions into (\ref{e130}) and (\ref{e131}) one finds,
\begin{eqnarray}
\label{e134}
\rho'' + (2 \nu_1 + \nu_2)\sinh\rho \cosh\rho &=&0\\
\xi'' + (\nu_3 +\frac{\nu_4}{2} )\sin(2\xi)&=&0.
\label{e135}
\end{eqnarray}

The corresponding solutions are given by,
\begin{eqnarray}
\rho'(\sigma^1)&=&\sqrt{C_1^2 - (2\nu_1 + \nu_2)\cosh^2\rho}\\
\xi' (\sigma^1)&=&\sqrt{C^2_2 - (2\nu_3 + \nu_4)\sin^2\xi}.
\end{eqnarray}
Following our previous discussions, the integration constants $ C_1 $ and $ C_2 $ may be estimated in terms of the turning points in the bulk geometry.

These result in the dispersion relation of the following form,
\begin{eqnarray}
H - Q=\frac{J}{\pi}\int_0^{\rho_+}d\rho \frac{C_1^2 - 2(2\nu_1 + \nu_2)\sinh^2\rho}{\sqrt{C_1^2 - (2\nu_1 + \nu_2)\cosh^2\rho}}\nonumber\\
+\frac{J}{2 \pi}\int_0^{\xi_{max}}d\xi \frac{C^2_2 + \nu_3 -2(2\nu_3 +\nu_4)\sin^2\xi}{\sqrt{C^2_2 - (2\nu_3 + \nu_4)\sin^2\xi}}
\label{e138}
\end{eqnarray}
where $ \rho_+ $ is the corresponding turning point(s) satisfying, $ \sinh^{2}\rho_+ = \frac{C^2_1}{2\nu_1 + \nu_2} $.

Evaluating the integrals above in (\ref{e138}) we find,
\begin{eqnarray}
\label{eqn68}
H-Q= \frac{J}{\pi}\left(2\sqrt{2\nu_1 +\nu_2}(\cosh\rho_+ -1)+ \frac{f(\xi_{max})}{2\sqrt{2\nu_3 + \nu_4}}\right) 
\end{eqnarray}
where we set the constants, $ C^2_1 = 2\nu_1 + \nu_2 $, $ C^2_2 = 2\nu_3 + \nu_4 $ and define the function,
\begin{eqnarray}
f(\xi_{max})=(\nu_3 +\nu_4) \log \left(\frac{\cos \left(\frac{\xi_{max} }{2}\right)-\sin \left(\frac{\xi_{max} }{2}\right)}{\sin \left(\frac{\xi_{max} }{2}\right)+\cos \left(\frac{\xi_{max} }{2}\right)}\right)+2(2\nu_3 + \nu_4)\sin\xi_{max}.
\end{eqnarray}
\subsection{A qualitative discussion on different SMT limits}
This Section is more dedicated towards exploring various SMT string limits in a qualitative way. In particular, we focus on the behavior of the spectrum $ \frac{(H-Q)}{J/4\pi} $ as function of the angular frequency of the string along the isometry direction of the target spacetime. 

\begin{figure}
\includegraphics[scale=.55]{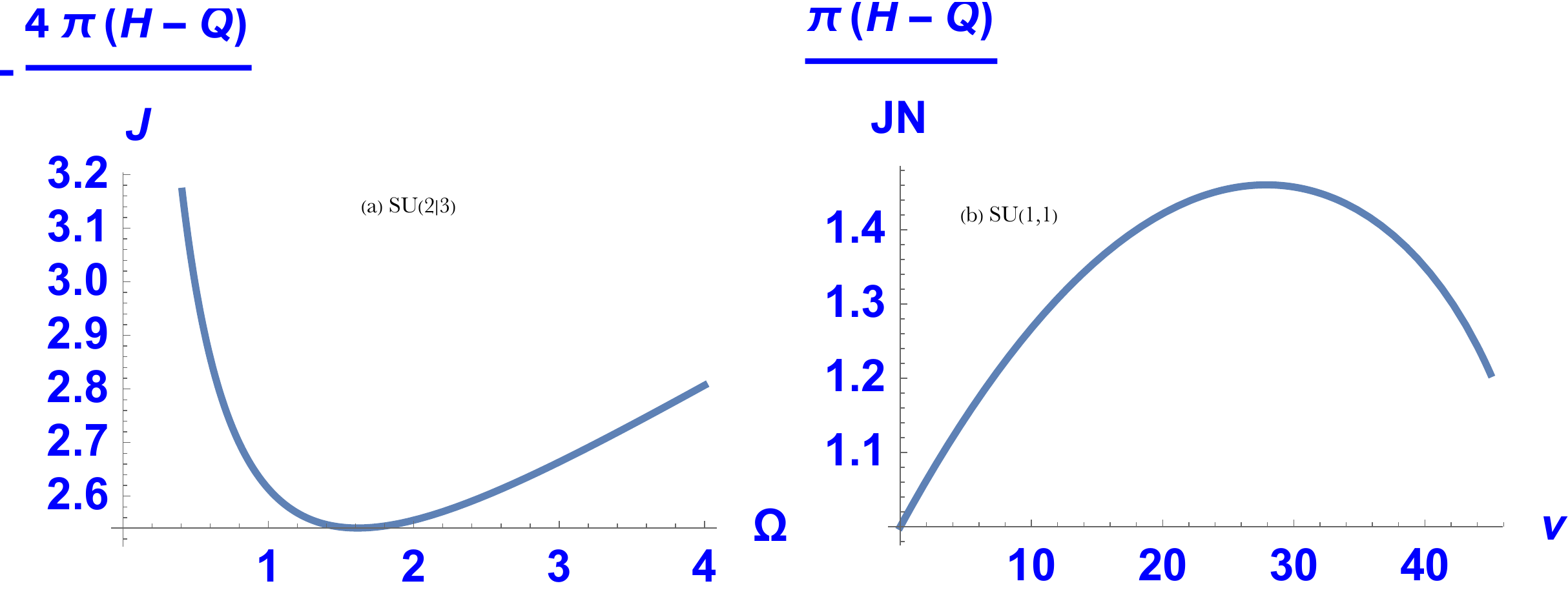}
  \caption{We plot the (regularised) difference $ H-Q $ against the angular frequency of the string corresponding to different SMT limits. (a) $H-Q  $ is plotted against the angular frequency ($ \Omega $) for $ SU(2|3) $ SMT strings, (b) $H-Q  $ is plotted against the angular frequency ($ \nu $) for short strings in the $ SU(1,1) $ SMT limit.}\label{smt}
\end{figure}

We start by discussing $ SU(2|3) $ SMT strings. The discussion for $ SU(2) $ subsector follows quite identically which we therefore prefer not to repeat here. As (\ref{eqn19}) suggests, the difference $ (H-Q) \sim \sqrt{\Omega} $ in the large frequency regime of the string. On the other hand, in the small frequency regime, it scales as $ \sim \frac{1}{\sqrt{\Omega}} $. In order to cure this divergence in the limit, $ \Omega \rightarrow 0 $ one has to add a counter term $ \sim \frac{bJ}{4\pi \sqrt{\epsilon}} $ to (\ref{eqn19}). Finally, the plot Fig.\ref{smt}a reveals the existence of a critical frequency which minimizes the difference $ (H-Q) $.

The $ SU(1,1) $ SMT dispersion relation (\ref{eqn33}), on the other hand, grows as $ \sim \sqrt{\nu} $ in the long string limit. However, it looks quite interesting in the short string limit (\ref{eqn39}), where the difference $ H-Q $ starts increasing with the increase in frequency ($ \nu $), thereby reaching a maximum at some critical value and then starts falling with the increase in frequency (Fig.\ref{smt}b). Qualitatively, this is just opposite to what we observe for $ SU(2|3) $ SMT strings. 

For $ SU(1,1|2) $ SMT sector (\ref{eqn46}), the difference $ H-Q $ depends on the relative value of the frequencies $ \nu_1 $ and $ \nu_2 $. For $ \nu_1 \gg \nu_2 $, this difference grows as $ \sqrt{\nu_1} $. On the other hand, for $ \nu_1 \ll \nu_2 $ the difference grows linearly ($ \sim \nu_2 $) with the frequency. 

On a similar note, we observe that for $ SU(1,2|2) $ sector (\ref{eqn53}), the difference ($ H-Q \sim (2\nu_1 + \nu_2 -\gamma^2)^3 $) grows as a cubic power of the frequency of rotation. Finally, for the $ PSU(1,2|3) $ strings (\ref{eqn68}), one recovers the square root behavior of the dispersion relation.
\section{Summary and final remarks}
The present paper provides an extensive analysis of different Spin-Matrix limits (SMT) proposed in the recent years. We explore various near BPS coreners of $ \mathcal{N} =4$ SYM by probing a class of $ U(1) $ Galilean geometries with semiclassical strings.

One should think of these SMT string solutions as a limiting case of rotating string configurations in $ AdS_5 \times S^5 $ \cite{Tseytlin:2010jv}-\cite{Tseytlin:2004xa}. The first step of this limiting proccedure includes taking a null reduction of $ AdS_5 \times S^5 $ string sigma models which is further accompanied by a (nonrelativistic) $ 1/c $ expansion of the world-sheet degrees of freedom. As a result of this double scaling limit, unlike the $ AdS_5 \times S^5 $ strings, the resulting string sigma model satisfies the Galilean conformal algebra \cite{Harmark:2017rpg}-\cite{Harmark:2018cdl}. Combining all these facts together, one should therefore think of the SMT string solutions as a nonrelativistic (double scaling) limit of $ AdS_5 \times S^5 $ relativistic string solutions those were studied in \cite{Tseytlin:2010jv}-\cite{Tseytlin:2004xa}. 

We obtain dispersion relations which may be typically expressed in the form, $ H-Q \sim f( \nu_i, N_i ) $. Here $ \nu_i $ is the frequency and $ N_i $ is the corresponding winding number of the string along the $ i $th isometry direction of the $ U(1) $ Galilean target space. 

Our analysis unfolds several interesting solutions associated with nonrelativistic sigma models which include both singular (spikes) as well as regular (magnons) solitonic configurations. Any systematic verification of these results from dual Spin-Matrix theory counterpart should be an exciting project to look for. 

Recently, there has been some progress along this line of investigation which involves a systematic investigation of Spin-Matrix theories with spin groups containing non compact $SU(1,1)$ factor \cite{Harmark:2019zkn}-\cite{Baiguera:2020mgk}. A similar control over other SMT limits at strong coupling and a verification of the corresponding string theory results should be something worthwhile to look for in the future.\\ \\ 
{\bf {Acknowledgements :}}
 The author is indebted to the authorities of IIT Roorkee for their unconditional support towards researches in basic sciences. The author would like to acknowledge The Royal Society, UK for financial assistance. The author would also like to acknowledge the Grant (No. SRG/2020/000088) received from The Science and Engineering Research Board (SERB), India.


\begin{thebibliography}{99}
\bibitem{Harmark:2006di}
T.~Harmark and M.~Orselli,
``Quantum mechanical sectors in thermal N=4 super Yang-Mills on R x S**3,''
Nucl. Phys. B \textbf{757}, 117-145 (2006)
doi:10.1016/j.nuclphysb.2006.08.022
[arXiv:hep-th/0605234 [hep-th]].

\bibitem{Harmark:2007px}
T.~Harmark, K.~R.~Kristjansson and M.~Orselli,
``Decoupling limits of N=4 super Yang-Mills on R x S**3,''
JHEP \textbf{09}, 115 (2007)
doi:10.1088/1126-6708/2007/09/115
[arXiv:0707.1621 [hep-th]].

\bibitem{Harmark:2008gm}
T.~Harmark, K.~R.~Kristjansson and M.~Orselli,
``Matching gauge theory and string theory in a decoupling limit of AdS/CFT,''
JHEP \textbf{02}, 027 (2009)
doi:10.1088/1126-6708/2009/02/027
[arXiv:0806.3370 [hep-th]].

\bibitem{Harmark:2014mpa}
T.~Harmark and M.~Orselli,
``Spin Matrix Theory: A quantum mechanical model of the AdS/CFT correspondence,''
JHEP \textbf{11}, 134 (2014)
doi:10.1007/JHEP11(2014)134
[arXiv:1409.4417 [hep-th]].

\bibitem{Harmark:2019zkn}
T.~Harmark and N.~Wintergerst,
``Nonrelativistic Corners of ${\cal N} = 4$ Supersymmetric Yang--Mills Theory,''
Phys. Rev. Lett. \textbf{124}, no.17, 171602 (2020)
doi:10.1103/PhysRevLett.124.171602
[arXiv:1912.05554 [hep-th]].

\bibitem{Baiguera:2020mgk}
S.~Baiguera, T.~Harmark, Y.~Lei and N.~Wintergerst,
``Symmetry structure of the interactions in near-BPS corners of $ \mathcal{N} = 4$ super-Yang-Mills,''
[arXiv:2012.08532 [hep-th]].

\bibitem{Gomis:2000bd}
J.~Gomis and H.~Ooguri,
``Nonrelativistic closed string theory,''
J. Math. Phys. \textbf{42}, 3127-3151 (2001)
doi:10.1063/1.1372697
[arXiv:hep-th/0009181 [hep-th]].

\bibitem{Gomis:2005pg}
J.~Gomis, J.~Gomis and K.~Kamimura,
``Non-relativistic superstrings: A New soluble sector of AdS(5) x S**5,''
JHEP \textbf{12}, 024 (2005)
doi:10.1088/1126-6708/2005/12/024
[arXiv:hep-th/0507036 [hep-th]].

\bibitem{Bergshoeff:2018yvt}
E.~Bergshoeff, J.~Gomis and Z.~Yan,
``Nonrelativistic String Theory and T-Duality,''
JHEP \textbf{11}, 133 (2018)
doi:10.1007/JHEP11(2018)133
[arXiv:1806.06071 [hep-th]].

\bibitem{Gomis:2019zyu}
J.~Gomis, J.~Oh and Z.~Yan,
``Nonrelativistic String Theory in Background Fields,''
JHEP \textbf{10}, 101 (2019)
doi:10.1007/JHEP10(2019)101
[arXiv:1905.07315 [hep-th]].

\bibitem{Bergshoeff:2019pij}
E.~A.~Bergshoeff, J.~Gomis, J.~Rosseel, C.~Simsek and Z.~Yan,
``String Theory and String Newton-Cartan Geometry,''
J. Phys. A \textbf{53}, no.1, 014001 (2020)
doi:10.1088/1751-8121/ab56e9
[arXiv:1907.10668 [hep-th]].

\bibitem{Harmark:2017rpg}
T.~Harmark, J.~Hartong and N.~A.~Obers,
``Nonrelativistic strings and limits of the AdS/CFT correspondence,''
Phys. Rev. D \textbf{96}, no.8, 086019 (2017)
doi:10.1103/PhysRevD.96.086019
[arXiv:1705.03535 [hep-th]].

\bibitem{Harmark:2018cdl}
T.~Harmark, J.~Hartong, L.~Menculini, N.~A.~Obers and Z.~Yan,
``Strings with Non-Relativistic Conformal Symmetry and Limits of the AdS/CFT Correspondence,''
JHEP \textbf{11}, 190 (2018)
doi:10.1007/JHEP11(2018)190
[arXiv:1810.05560 [hep-th]].

\bibitem{Grosvenor:2017dfs}
K.~T.~Grosvenor, J.~Hartong, C.~Keeler and N.~A.~Obers,
``Homogeneous Nonrelativistic Geometries as Coset Spaces,''
Class. Quant. Grav. \textbf{35}, no.17, 175007 (2018)
doi:10.1088/1361-6382/aad0f9
[arXiv:1712.03980 [hep-th]].

\bibitem{Hansen:2018ofj}
D.~Hansen, J.~Hartong and N.~A.~Obers,
``Action Principle for Newtonian Gravity,''
Phys. Rev. Lett. \textbf{122}, no.6, 061106 (2019)
doi:10.1103/PhysRevLett.122.061106
[arXiv:1807.04765 [hep-th]].

\bibitem{Christensen:2013lma}
M.~H.~Christensen, J.~Hartong, N.~A.~Obers and B.~Rollier,
``Torsional Newton-Cartan Geometry and Lifshitz Holography,''
Phys. Rev. D \textbf{89}, 061901 (2014)
doi:10.1103/PhysRevD.89.061901
[arXiv:1311.4794 [hep-th]].

\bibitem{Hartong:2014pma}
J.~Hartong, E.~Kiritsis and N.~A.~Obers,
``Schrödinger Invariance from Lifshitz Isometries in Holography and Field Theory,''
Phys. Rev. D \textbf{92}, 066003 (2015)
doi:10.1103/PhysRevD.92.066003
[arXiv:1409.1522 [hep-th]].

\bibitem{Hartong:2014oma}
J.~Hartong, E.~Kiritsis and N.~A.~Obers,
``Lifshitz space–times for Schrödinger holography,''
Phys. Lett. B \textbf{746}, 318-324 (2015)
doi:10.1016/j.physletb.2015.05.010
[arXiv:1409.1519 [hep-th]].

\bibitem{Roychowdhury:2020cnj}
D.~Roychowdhury,
``Nonrelativistic strings on $ R \times S^2 $ and integrable systems,''
Nucl. Phys. B \textbf{961}, 115220 (2020)
doi:10.1016/j.nuclphysb.2020.115220
[arXiv:2003.02613 [hep-th]].

\bibitem{Hartong:2015zia}
J.~Hartong and N.~A.~Obers,
``Hořava-Lifshitz gravity from dynamical Newton-Cartan geometry,''
JHEP \textbf{07}, 155 (2015)
doi:10.1007/JHEP07(2015)155
[arXiv:1504.07461 [hep-th]].

\bibitem{Roychowdhury:2020yun}
D.~Roychowdhury,
``Nonrelativistic giant magnons and $ SU(1,2|3) $ limit of strings in $ AdS_5 \times S^5 $,''
[arXiv:2010.05179 [hep-th]].

\bibitem{Christensen:2013rfa}
M.~H.~Christensen, J.~Hartong, N.~A.~Obers and B.~Rollier,
``Boundary Stress-Energy Tensor and Newton-Cartan Geometry in Lifshitz Holography,''
JHEP \textbf{01}, 057 (2014)
doi:10.1007/JHEP01(2014)057
[arXiv:1311.6471 [hep-th]].

\bibitem{Roychowdhury:2019sfo}
D.~Roychowdhury,
``Semiclassical dynamics for torsional Newton-Cartan strings,''
doi:10.1016/j.nuclphysb.2020.115132
[arXiv:1911.10473 [hep-th]].

\bibitem{Gallegos:2019icg}
A.~D.~Gallegos, U.~G\"ursoy and N.~Zinnato,
``Torsional Newton Cartan gravity from non-relativistic strings,''
JHEP \textbf{09}, 172 (2020)
doi:10.1007/JHEP09(2020)172
[arXiv:1906.01607 [hep-th]].

\bibitem{Roychowdhury:2019olt}
D.~Roychowdhury,
``Nonrelativistic pulsating strings,''
JHEP \textbf{09}, 002 (2019)
doi:10.1007/JHEP09(2019)002
[arXiv:1907.00584 [hep-th]].

\bibitem{Harmark:2019upf}
T.~Harmark, J.~Hartong, L.~Menculini, N.~A.~Obers and G.~Oling,
``Relating non-relativistic string theories,''
JHEP \textbf{11}, 071 (2019)
doi:10.1007/JHEP11(2019)071
[arXiv:1907.01663 [hep-th]].

\bibitem{Roychowdhury:2020dke}
D.~Roychowdhury,
``Nonrelativistic spinning strings,''
JHEP \textbf{11}, 044 (2020)
doi:10.1007/JHEP11(2020)044
[arXiv:2008.08895 [hep-th]].

\bibitem{Harmark:2020vll}
T.~Harmark, J.~Hartong, N.~A.~Obers and G.~Oling,
``Spin Matrix Theory String Backgrounds and Penrose Limits of AdS/CFT,''
[arXiv:2011.02539 [hep-th]].

\bibitem{Roychowdhury:2020kma}
D.~Roychowdhury,
``Nonrelativistic giant magnons from Newton Cartan strings,''
JHEP \textbf{02}, 109 (2020)
doi:10.1007/JHEP02(2020)109
[arXiv:2001.01061 [hep-th]].

\bibitem{Tseytlin:2010jv}
A.~A.~Tseytlin,
``Review of AdS/CFT Integrability, Chapter II.1: Classical AdS5xS5 string solutions,''
Lett. Math. Phys. \textbf{99}, 103-125 (2012)
doi:10.1007/s11005-011-0466-0
[arXiv:1012.3986 [hep-th]].

\bibitem{Tseytlin:2004xa}
A.~A.~Tseytlin,
``Semiclassical strings and AdS/CFT,''
[arXiv:hep-th/0409296 [hep-th]].

\end{thebibliography}
\end{document}